\documentclass[10pt]{iopart}
\usepackage{epsfig}
\usepackage{amstext}
\usepackage{amssymb}
\usepackage{amsfonts}
\usepackage{graphicx}
\usepackage{axodraw}
\usepackage{vmargin}
\usepackage{amscd}
\usepackage{bbold}
\usepackage{mathrsfs}
\usepackage{fancyheadings}
\usepackage{pifont}
\usepackage{pst-node}

\newcommand{\kp}{K^+}

\newcommand{\NPA}{Nucl.\ Phys.\ A}

\newcommand{\PRC}{Phys.\ Rev.\ C}

\newcommand{\be}{\begin{equation}}
\newcommand{\ee}{\end{equation}}
\begin{document}
\title{Transport Theories for Heavy Ion Collisions in the 1 AGeV Regime}
\author{E.E. Kolomeitsev$^1$, C. Hartnack$^2$, H.W. Barz$^3$, M. Bleicher$^4$,
E. Bratkovskaya$^4$, W. Cassing$^5$, L.W. Chen$^{6,7}$, P. Danielewicz$^8$, 
C. Fuchs$^9$, 
T. Gaitanos$^{10}$, C.M. Ko$^6$, A. Larionov$^{5\dagger}$,  M. Reiter$^4$,
Gy. Wolf$^{11}$, J. Aichelin$^2$\footnote{invited speaker}}
\address{$^1$ The Niels Bohr Institute, Blegdamsvej 17, DK-2100, Copenhagen, Denmark,\\
School of Physics  and Astronomy, University of Minnesota,
Minneapolis, MN-55455, USA }
\address{$^2$ SUBATECH,
University of Nantes - IN2P3/CNRS - EMN, F-44072 Nantes, France}
\address{$^3$ Forschungszentrum Rossendorf, D-01314 Dresden, Germany}
\address{$^4$ Institut f\"ur Theoretische Physik der Wolfgang Goethe
Universit\"at, D-60054 Frankfurt, Germany}
\address{$^5$ Institut f\"ur Theoretische Physik der
Universit\"at Giessen, D-35392 Giessen, Germany}
\address{$^6$ Cyclotron Institute and Physics Department, 
Texas A\&M University, College Station, TX-77843, USA  }
\address{$^{7}$ Department of Physics, Shanghai Jiao Tong University,
Shanghai 20030, China}
\address{$^8$ National Superconducting Cyclotron Laboratory, Michigan State 
University, East Lansing, MI-48824, USA }
\address{$^9$ Institut f\"ur Theoretische Physik Universit\"at T\"ubingen, 
D-72076 T\"ubingen, Germany}
\address{$^{10}$ Laboratori Nazionali del Sud INFN, I-95123 Catania, Italy}
\address{$^{11}$ KFKI Budapest, POB 49, H-1525 Budapest, Hungary}

\address{$^\dagger$ on leave from I.V. Kurtchatov Institute,  123182 Moscow,Russia}
\begin{abstract}

We compare multiplicities as well as rapidity and transverse momentum 
distributions of protons, pions and kaons calculated within presently available transport 
approaches for heavy ion collisions around 1 AGeV. For this purpose, three reactions have been 
selected: Au+Au at 1 and 1.48 AGeV and Ni+Ni at 1.93 AGeV.  
\end{abstract}
\date{\today}
\section{Introduction}
Heavy ion collisions in the energy range from 50 AMeV to  several AGeV are quite
complex. The single particle spectra are not isotropic and the anisotropy 
depends significantly
on the centrality of the reaction. With increasing energy one observes the rise
and fall of multifragmentation, where multifragmentation means that in each 
single collision a multitude of intermediate mass fragments is produced. 
While above $E_{kin}=289(1583)MeV$, $\pi$'s($K^+$'s) can be produced in elementary
nucleon-nucleon collisions, experiments have shown that in collisions 
between nuclei Fermi motion and multiparticle effects allow for meson production at energies 
well below the threshold in an elementary reaction.
Finally, nucleons can be excited to resonances which - according to observations
in $\gamma$A  
reactions - change their properties in a nuclear environment. 

This complexity of the reactions makes it very difficult to link the 
experimental results in a unique way with the specific underlying physical processes.
The challenge to interpret the experimental results was first met by Yariv and
Fr\"ankel \cite{yar} and Cugnon \cite{cug} who took advantage of the increasing computer power at
that time and developed the so-called "cascade approach" in which nucleons
behave similarly to classical billiard balls and scatter with the free elementary cross
sections. This computational approach allowed for  completely new 
insights into the reaction mechanism and allowed for the first time to interpret
the nonequilibrium features of the observed spectra.  Disregarding binding energy
and mean field effects, these approaches have been limited to small systems at higher energies.
The extension to lower energies became possible half a decade later
with the numerical realization of the
Boltzmann-Uehling- Uhlenbeck (BUU) \cite{aic} or Vlasov-Uehling- Uhlenbeck (VUU)
\cite{sto} approaches which supplemented cascade approaches by the
introduction of an attractive mean field and of Pauli blocking, which suppresses
collisions if the phase space is already occupied by other nucleons. 
This made it possible to describe heavy ion reactions down to energies of 20
AMeV.
Over the years these approaches became more and more refined as a result
of constructive competitions among different
groups and many physical questions have been addressed and answered.
The pre-equilibrium proton emission has been quantitatively reproduced.
It is nowadays established that subthreshold mesons like $\eta$, $K$ or $\rho$ 
are created in elementary collisions, where at least one of the collision 
partners had gained additional energy in previous collisions. It is further
established
that resonances are the primary source for the produced pions. 
 
In the meantime, the simulation programs have matured to the level which allows 
to go even beyond the original goal of describing the elementary reaction features: 
They are nowadays commonly used as
a tool to find out whether the experimental results allow for a determination
of physical quantities for which no solid theoretical predictions are available.
These quantities include the energy which is needed to compress
nuclear matter as well as the properties of resonances and mesons in a
hadronic environment. 

For the following three reasons it seems to be useful to compare the simulation programs
which have been developed over the years: 

1) The simulation programs are rather complex - consisting of several 
thousand program lines - and  use quite different numerical techniques. Thus, 
one has to assess whether the different numerical procedures lead to the same 
results.

2) All programs rely on inputs. Those inputs include all the needed elementary 
cross sections. Among those there are cross sections which 
are not known experimentally (like those including a baryonic resonance in 
the entrance channel), and different theoretical implementations have been 
introduced. It is evident that different elementary cross sections can yield 
different results.

3) In all programs resonances are produced. Their properties in the hadronic
environment as well as their time evolution are little known. The programs use
different parameterizations for this and it is not a priori evident how those
differences influence the final result.

Thus the detailed comparison presented in the present paper 
may serve as the first step towards a
critical assessment of the predictive power of the simulation programs and to 
identify the features which have to be improved.
We would like to mention that some elements towards a future development are
already available. Very detailed calculations of the spectral function of kaons
in infinite matter have been published \cite{lut} and first
steps towards describing the time evolution of resonances in matter have been 
made \cite{cas2,leu,kno}. These advances, based on the gradient expansion
of the Kadanoff-Baym equation, are difficult to formulate for test particles,
and further approximations have been shown to be necessary for actual applications
\cite{cas2}. Apart from these approaches there are a couple of studies providing
information on the off-shell transition rates \cite{fu1,tol}.

\section{The simulation programs}
In this comparison nearly all the presently available simulation programs have 
taken part:
The simulation program from the Budapest/Rossendorf group \cite{ros}, the HSD program developed
in Giessen by Bratkovskaya and Cassing \cite{cc1,hsd}, the simulation program
developed by Bratkovskaya, Effenberger, Larionov and Mosel 
in Giessen \cite{gim}, the code developed by 
Danielewicz in Michigan \cite{dan}, the RVUU approaches of the 
Texas A\&M group \cite{chen} and the Munich/Catania/T\"ubingen group \cite{gai}.
Besides these true BUU or VUU models, there are several models which are based
on the Quantum Molecular Dynamics approach. These n-body approaches allow for
the description of fragment production but as far as the observables 
discussed here they should give the same results. Therefore, we just refer
to ref. \cite{joerg} for the general features and do not discuss the differences here. 
These simulation models include the IQMD \cite{iqmd} and URQMD \cite{urqmd}, both developed in a 
collaboration between Nantes and Frankfurt, as well as the QMD of the 
T\"ubingen group (by Fuchs)\cite{fuc}. These programs contain a different
number of baryonic resonances as can be inferred from table \ref{width}.
The URQMD code has been developed for higher (SPS to RHIC) energies and 
includes neither a binding energy for nucleons nor a kaon-nucleon potential
but all known mesonic and baryonic resonances. Also its production mechanism 
for $K^+$ is different from other approaches. Therefore it is only included
to illuminate the differences.    
 
\begin{table}[hbt]
\centering
\begin{tabular}{|c|c|c|} \hline
Program  &$\tau$&Resonances\\ \hline
 Barz  & $\tau_{iso}$&$\Delta's$,N's with M $<$ 2 GeV   \\ \hline
 RVUU (Chen) & $\tau_{iso}$&$\Delta(1232)$  \\ \hline
 Danielewicz & $\tau_{iso}$&$\Delta(1232),N^*(1440)$ \\ \hline
 QMD (Fuchs) & $\tau_{iso}$&$\Delta(1232),N^*(1440)$  \\ \hline
 Gaitanos & $\tau_{iso}$&$\Delta(1232),N^*(1440)$  \\ \hline
 HSD (Cassing) & $\tau_{iso}$&$\Delta(1232),N^*(1440),N(1535)$  \\ \hline
 IQMD (Hartnack)& $\tau_{wig}$&$\Delta(1232)$  \\ \hline
 Larionov& 1/120 MeV&$\Delta's,N's$ with M $<$ 2 GeV  \\ \hline
\end{tabular}
\caption{The width of the $\Delta$ resonance and the resonances 
employed in the programs.}
\label{width}
\end{table}  \section{The Procedure}

The simulations programs, which are compared in this article, have been 
frequently compared with experimental data and usually gave quite
satisfying agreement. It is, however, difficult to base program comparisons 
on those calculations because rarely the same data have been compared  with different simulation
programs and in the rare cases one has done so, it is not evident that the
same experimental filter or the same centrality selection has been employed. 
To assess the predictive power of these programs and to see whether 
differences in the predictions are sufficiently large to be experimentally 
relevant, it is much better to compare the results of the programs directly without any cuts.
In addition, all programs employ the impact parameter as an input variable
and therefore it is most convenient to compare the results at a given impact
parameter. This excludes of course any comparison with experiment. In this
publication we concentrate on 3 reactions: Au + Au at 1 and 1.48 AGeV and Ni+Ni
at 1.93 AGeV, all at an impact parameter of $b$ = 1 fm as well as the pion yield as
a function of the impact parameter. This choice was guided by the available
experimental results.

\section{Protons}
\begin{figure}
\includegraphics[width=15cm]{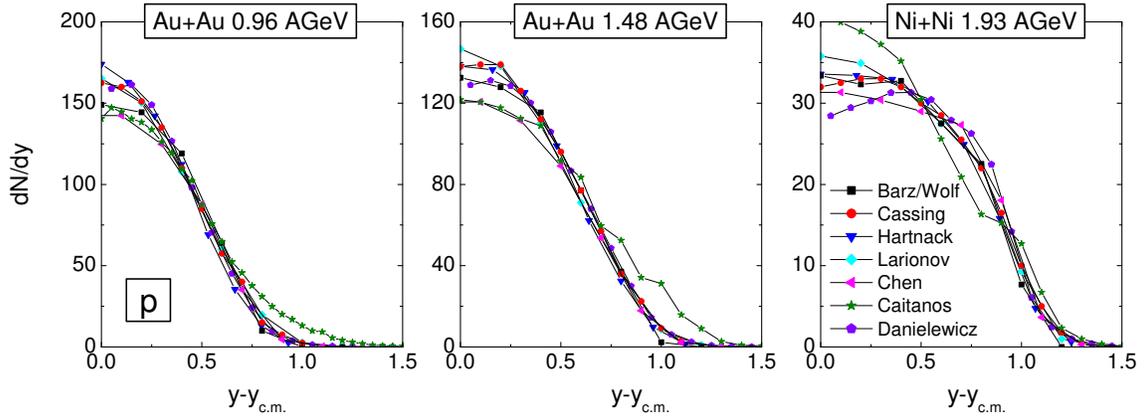}
\caption{Final proton rapidity distribution at $b$ = 1fm in the different approaches}
\label{proton-y}
\end{figure}

\begin{figure}
\includegraphics[width=15cm]{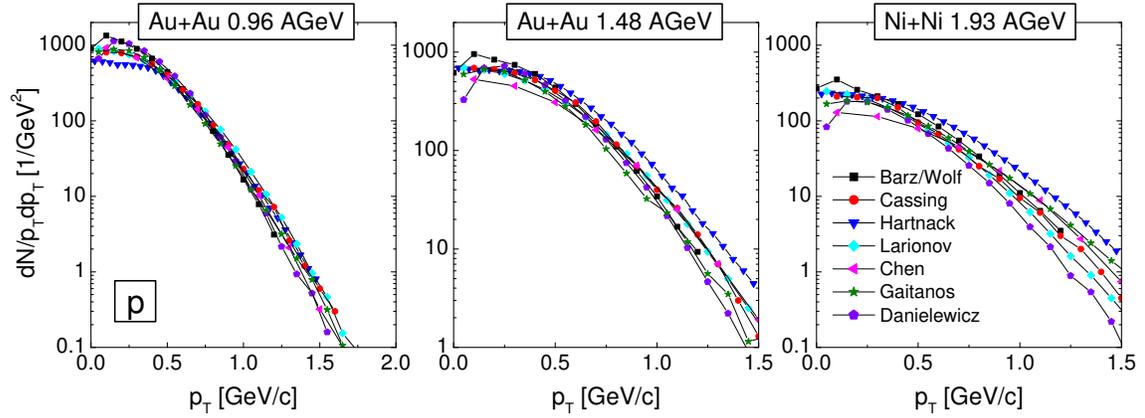}
\caption{Final  proton transverse momentum distribution at $b$ = 1fm in the different approaches}
\label{proton-pt}
\end{figure}

The final proton rapidity distribution measures the amount of stopping and 
reflects how much energy becomes available for particle production. The recent 
analysis of the
FOPI collaboration \cite{fopi} shows that, in the center of mass 
system, the variance of the transverse rapidity distribution is always 
smaller than twice the variance of the longitudinal rapidity distribution
even in the most central collisions. If the system were in thermal equilibrium 
one would expect that the ratio R of the variances equals one. Experimentally
the largest degree of thermalisation (R $\approx$ 0.85) is observed  
for central Au + Au collisions around 400 AMeV. Below that energy Pauli blocking
is still too active, while above the NN cross section becomes more and more forward 
peaked and therefore the energy transfer into the perpendicular directions
is lowered. In addition, more energy has to be transfered in order 
to come to equilibrium.
The rapidity and transverse momentum distributions from simulations, figs.\ref{proton-y} 
and \ref{proton-pt}, reflect this approach to equilibrium. 
For the heavy systems the final rapidity distributions is peaked at or
very close to midrapidity and the simulations differ only little. Initially
the protons are located around $y_{cm} = 0.68, 0.8, 0.89$ for the
three reactions, respectively. The resulting rapidity shift for the large systems is thus large.
A slightly smaller shift is observed for the light system and there also differences develop 
between the predictions of the different programs. This kind of differences should be measurable 
and can help to validate the assumptions made. In transverse direction and on logarithmic scale,
we find a general agreement among the programs for Au+Au collisions. However, 
in the lighter system pronounced differences develop. They seem to be 
correlated with the rapidity distribution and reflect a different degree of stopping 
in the different approaches. The differences are most pronounced at small transverse momenta. 
To fill this part of phase space many collisions are necessary.
Therefore its form is a measure for the number of collisions during the
interaction. Table \ref{provar} summarizes the average squared momenta in the center of mass system
at the end of the reaction, along the beam (z) direction and in the direction (y)
perpendicular to the reaction plane. 
\begin{table}[hbt]
\centering
\begin{tabular}{|c|c|c|c|c|c|c|c|c|c|} \hline
&\multicolumn{3}{c|}{1 AGeV Au+Au}&\multicolumn{3}{c|}{1.48 AGeV Au+Au}&
\multicolumn{3}{c|}{1.93 AGeV Ni+Ni}\\ \hline
Program & $\langle p_z^2 \rangle$ & $\langle  p_y^2 \rangle$&
$\frac{\langle p_y^2 \rangle}{\langle p_z^2 \rangle }$ 
& $\langle p_z^2 \rangle$ & $\langle  p_y^2  \rangle$&
$\frac{\langle p_y^2 \rangle}{\langle p_z^2 \rangle }$
& $\langle p_z^2 \rangle$ & $\langle  p_y^2  \rangle$&
$\frac{\langle p_y^2 \rangle}{\langle p_z^2 \rangle }$
\\ \hline
& \multicolumn{2}{c|}{$(GeV^2/c^2)$ } & &\multicolumn{2}{c|}{$(GeV^2/c^2)$ } && \multicolumn{2}{c|}{$(GeV^2/c^2)$ } &  \\ \hline
Barz/Wolf  &0.157 & 0.109&0.69 &0.230 &0.146 &0.64&0.351 & 0.147&0.42  \\ \hline
 RVUU (Chen) &0.165 &0.121 &0.73&0.256 &0.172 &0.67&0.416 &0.172 &0.42   \\ \hline
 Danielewicz &0.162 &0.102 &0.63&0.257 &0.126 &0.49& 0.407&0.104 &0.26 \\ \hline
 QMD (Fuchs) &0.158&0.111 &0.70 &0.262 &0.138 &0.53&0.413 &0.107 & 0.26\\ \hline
 Gaitanos& 0.131&0.116 &0.89 &0.240 &0.142 &0.60&0.300 &0.104 &0.35\\ \hline
 HSD (Cassing) &0.138 &0.120 &0.87  & 0.219&0.157 &0.72&0.361 &0.148 &0.41\\ \hline
 IQMD (Hartnack)& 0.143& 0.121& 0.85 & 0.226 &0.161 & 0.72 &0.378 &0.158 &0.42\\ \hline
 Larionov& 0.163&0.124 &0.76&0.252 &0.159 &0.63 &0.405 &0.136 & 0.34\\ \hline
\end{tabular}
\caption{Average out-of-plane and longitudinal squared momenta for protons in the different
reactions.}
\label{provar}
\end{table}
\section{Pions}
The direct pion production cross section in nuclear interactions is small. 
Therefore all pions are created in simulations via resonances. These resonances are
produced with the free inelastic NN collisions - which is well known - and in the
region of interest comparable to the elastic cross section. In our energy region
these are mostly $\Delta$ resonances which disintegrate into a $\pi$ and a
nucleon. The problem is how to treat this resonance in matter as is required for
the simulation programs.
The $\Delta$ resonance, as noted in the particle data booklet, has a width 
$\Gamma$ of about 120 MeV and hence a lifetime of $\tau = \frac{1}{\Gamma} = 1.7
fm/c$. Therefore it disintegrates inside the nuclear medium. The width has been 
determined by analyzing the phase shift in $\pi$N scattering as a function of
$\sqrt{s}$. Around $\delta (\sqrt{s}) = \pi/2$ the dependence on the energy may 
be approximated with $\delta (\sqrt{s})= tan^{-1} (\frac{\Gamma}
{2(m_{\Delta}-\sqrt{s})})$. The lifetime as inverse of a constant width $\Gamma$
is justified if a) the width were small as compared to the mass
difference between the resonance and its decay products 
($ \Gamma << m_\Delta -(m_N+m_\pi)$) and if b) the energy
spread of the particle wave packets were large compared to $\Gamma$. 
This is, however, not the case: the nucleons are represented by delta functions in
energy. Therefore one has to address the question which lifetime should be
assigned to the object which is created in an inelastic nucleon-nucleon 
collision with a sharp $\sqrt{s}$ and which yields two nucleons and a pion 
in the exit channel. There are two approaches in the
literature. Ericson and Weise \cite{eri}, in particular, have analyzed 
the $\pi N \Delta$ system
in a relativistic isobar model and found for the scattering amplitude 
\be
f^{\Delta}_{33} = \frac{\gamma {\bf k^2}}{m_\Delta^2 - s - i \gamma |{\bf
k}|^3},
\ee
a form which resembles the usual relativistic Breit-Wigner form and yields an
energy dependent $\Delta$ decay width of 
\be
\Gamma_\Delta (|{\bf k}|)= 2 \gamma |{\bf k}|^3=\frac{2f_\Delta ^2|{\bf k}|^3 
M_\Delta}{12 \pi m_\pi^2 \sqrt{s}}. 
\ee
With $f_\Delta^2 = 4.02$ this form reproduces very well the $\pi$ N data.
$\Gamma_\Delta (|\bf k|)$ decreases with decreasing $\sqrt{s}$
and therefore a lifetime  defined as  
\be
\tau_{iso} \propto \frac{1}
{\Gamma_\Delta (|\bf k|)}
\ee 
increases with decreasing $\sqrt{s}$.
 
\begin{figure}
\includegraphics[width=8cm]{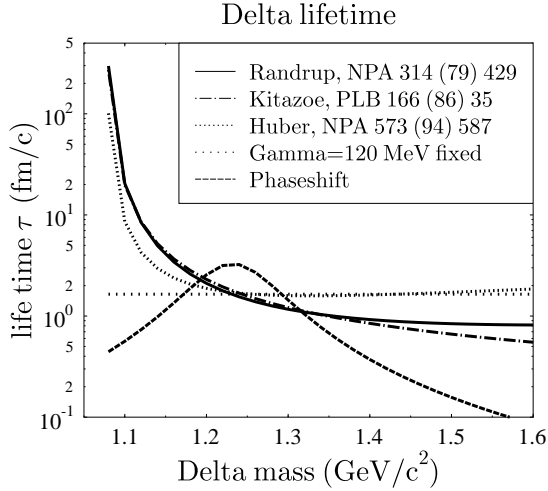}
\caption{The different $\Delta$ lifetimes which have been proposed.}
\label{lifetime}
\end{figure}

On the other hand, the lifetime of a resonant state has been already analyzed
by Wigner \cite{wig}. In recent time this considerations have been generalized
by Danielewicz and Pratt \cite{danlif}.
They found that quantum mechanically the reaction slows down with 
a time delay between the outgoing and the incoming wave of
\be
\tau_{wig} = 2\frac{ d \delta (E)}{dE}.    
\ee
Hence maximal time delay occurs at the center of the resonance. 
Fig. \ref{lifetime} show the $\Delta$ lifetimes in the different approaches
as a function of the $\Delta$ mass. Randrup, Kitazoe and Huber are synonymous 
for three rather similar approaches which calculate the width using 
the p-wave scattering amplitude as described above.
The description used in the standard versions of the 
different programs are shown in Table \ref{width}.

The influence of the
choice of $\tau$ on the final number of pions is by far not
negligible. Almost independent of energy and size of the system 
$\tau_{iso}$ may yield up to 60\% more pions than $\tau_{wig}$.
If the $\Delta$ has a short lifetime, several generations of $\pi$ and $\Delta$
are produced.
In an expanding system the average available center of mass energy in the NN 
system decreases and the cross section for the production of new
$\Delta$'s becomes smaller whereas for the backward reaction it is little influenced.
 
\begin{figure}[hbt]
\includegraphics[width=15cm]{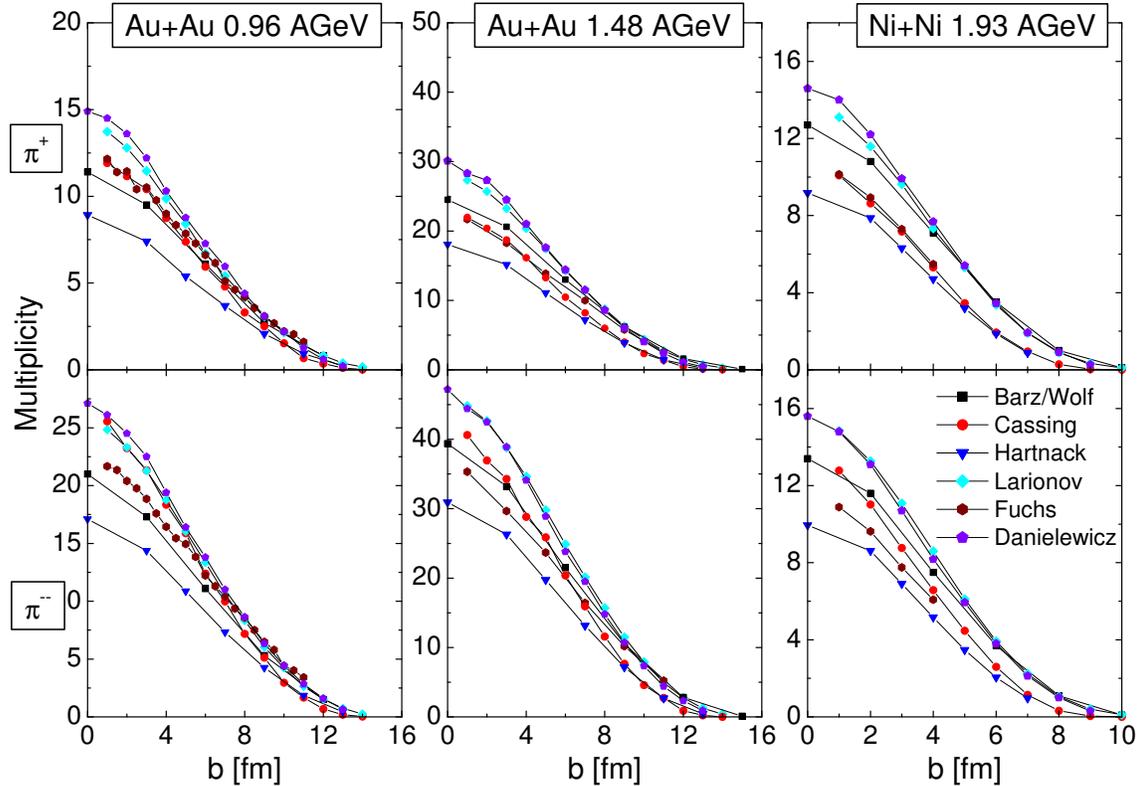}
\caption{Final $\pi^+$ and $\pi^-$ multiplicity as functions of the impact
parameter in the standard versions of the different approaches characterized
by different descriptions of the $\Delta$ lifetime (see table \ref{width}).}
\label{pion-b}
\end{figure}

\begin{figure}[hbt]
\includegraphics[width=15cm]{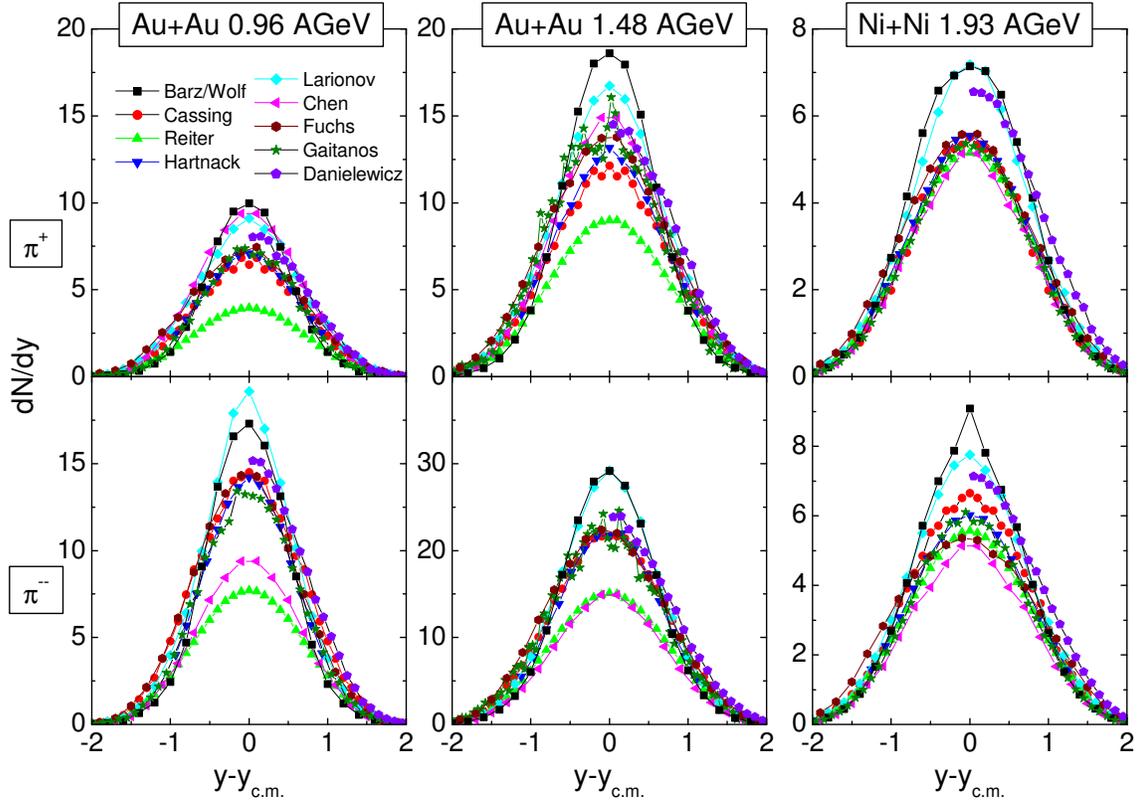}
\caption{Final $\pi^+$ and $\pi^-$ rapidity distributions at $b$ = 1fm in the different approaches 
with an enforced $\Delta$ lifetime of 1/120 MeV.}
\label{pi+-y}
\end{figure}

\begin{figure}
\includegraphics[width=15cm]{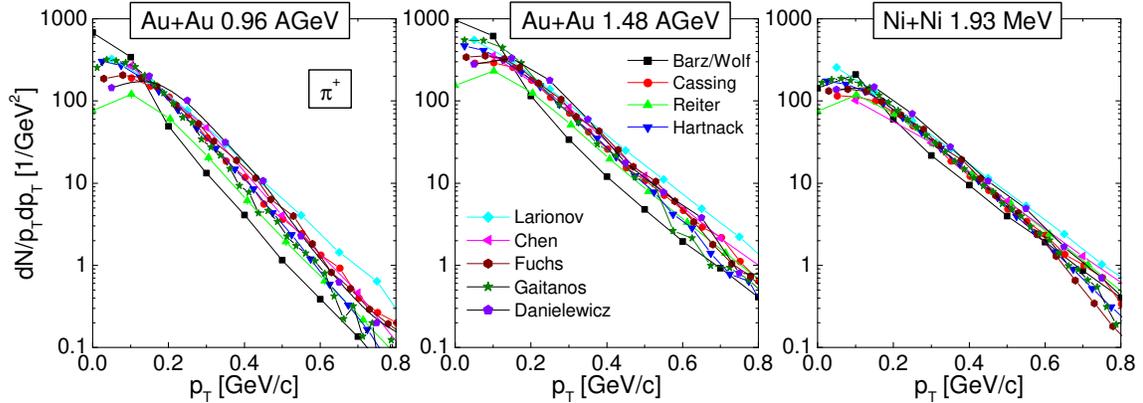}
\caption{Final $\pi^+$ transverse momentum distribution at $b$= 1 fm and $|y_{cm} < 0.5|$ 
in the different approaches with an enforced $\Delta$ lifetime of 1/120 MeV.}
\label{pion-pt}
\end{figure}

Fig. \ref{pion-b} shows the number of produced pions as a function of the impact
parameter for the three reactions. We see the expected significant variations of the
total yield. 
In order to put the results on an equal footing we have decided that all other
calculations presented here use a constant lifetime of $\tau = 1 /\Gamma, \Gamma =
120 \ MeV$.
The rapidity and transverse momentum distributions, obtained with this constant lifetime,
are displayed in figs. \ref{pi+-y} and \ref{pion-pt}, respectively.
We see, first of all, that the differences between the calculations are reduced
but not eliminated. The calculation of Reiter, based on URQMD, has to 
be interpreted with care because in this calculation neither potentials are present
nor - being an approach for high energies - special care has been taken to
model the inelastic cross section close to threshold. In the program of
Chen only the isospin averaged pion yield is available. For that program
we have therefore  plotted here for $\pi^-$ and $\pi^+$ one third of the total 
$\pi$ yield. Aside from Reiter's results, the rapidity distributions 
are rather similar for the majority of the
programs. There are, however, some
exceptions which show a narrower distribution. The experimental ratio 
$ R = \pi^- /\pi^+$ at 1 AGeV is close to the value R=1.95 \cite{sen} 
which one expects from the Clebsch Gordon coefficients if (in this 
neutron/proton asymmetric system)  the $\pi$'s are exclusively produced 
via $\Delta$ resonances. Most of the simulation programs reproduce this value.
The slopes of the transverse momentum spectra at midrapidity 
are very similar as well. The differences concern exclusively $\pi$'s with 
very small $p_T$ values. This part of the spectra is sensitive to $\Delta$'s 
with a very small mass.
We would like to stress that some of the programs include
higher ($N^*$ or $\Delta^*$) resonances
whereas others are limited to the $\Delta$ resonance. This may be of importance
at the higher beam energies.
Generally the simulation programs seem to overpredict the pion yield.
A compilation of the available data in ref. \cite{sen} shows an average total
pion yield per participant of $ \pi^+ + \pi^0 + \pi^- = 1.5(\pi^+ + \pi^-) $ 
= 0.075 and therefore about 30 pions in the most central collisions of
Au+Au at 1 AGeV.
Please note that the slope of the pion transverse momentum spectra is rather
different from that of the protons. This is due the convolution of
the $\Delta$ transverse momentum distribution with its the decay kinematics.
\section{$K^+$ Mesons}
\begin{figure}[hbt]
\includegraphics[width=15cm]{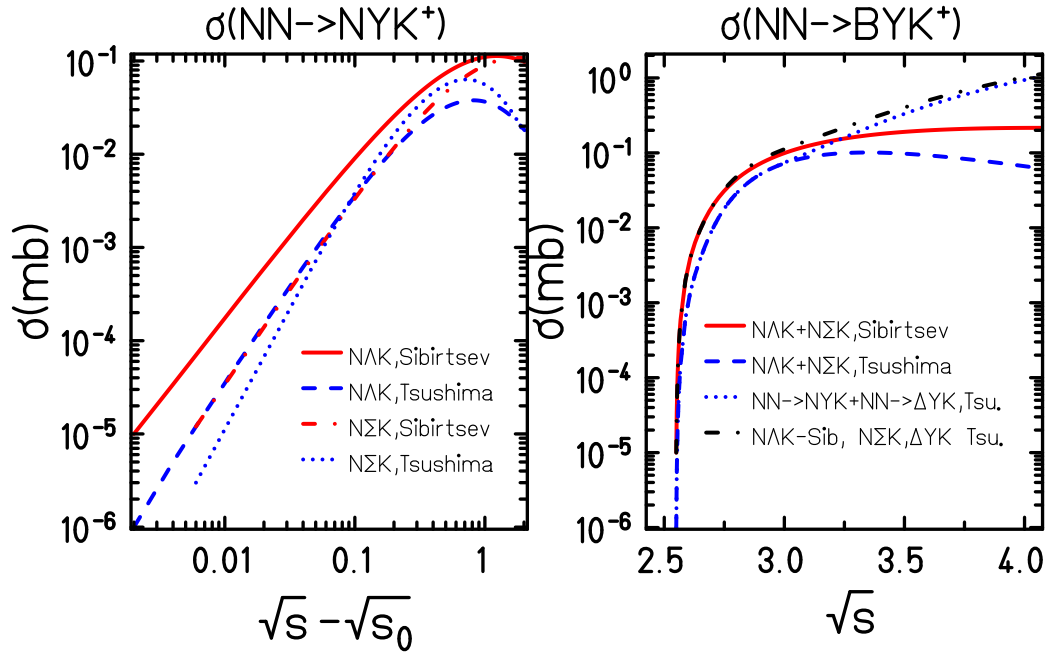}
\caption{The isospin averaged cross sections $NN\rightarrow K^+$.}
\label{kp-cross}
\end{figure}

\begin{figure}[hbt]
\includegraphics[width=15cm]{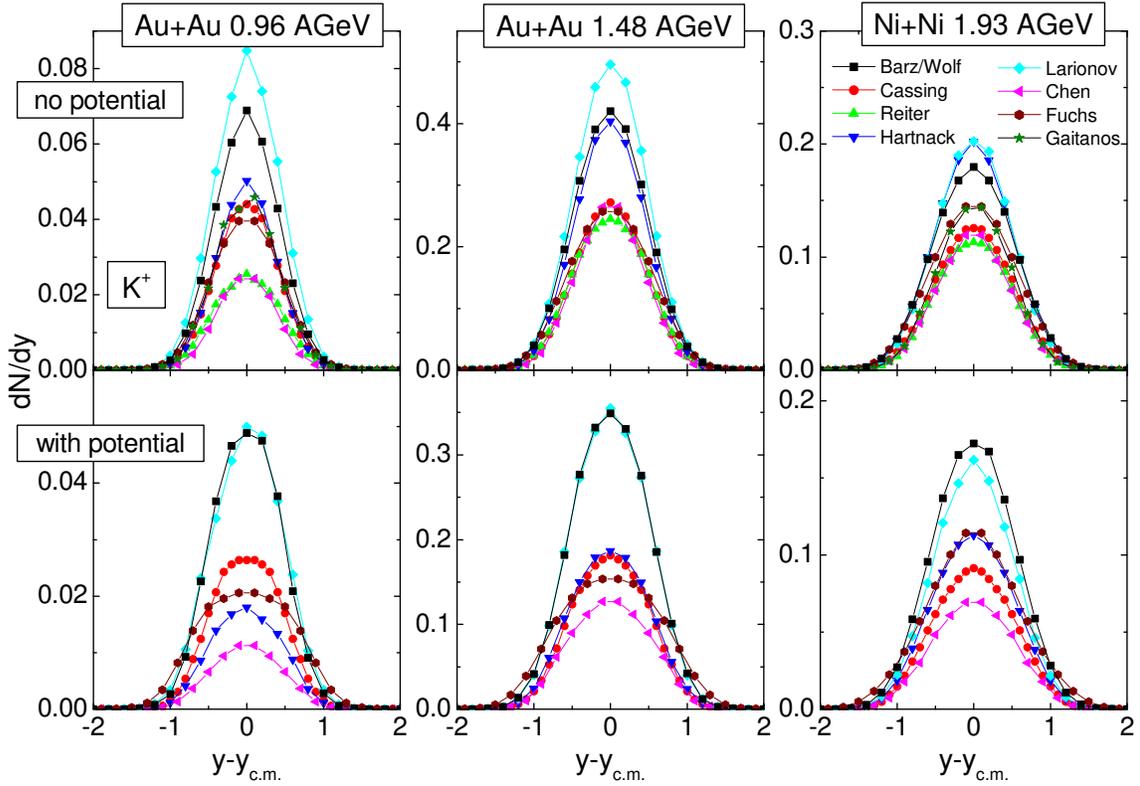}
\caption{Final $K^+$ rapidity distribution at $b$ = 1 fm and with an enforced $\Delta$ lifetime of 
1/120 MeV (top row without, bottom row with KN potential) in the different approaches.}
\label{kp-y}
\end{figure}

\begin{figure}[hbt]
\includegraphics[width=15cm]{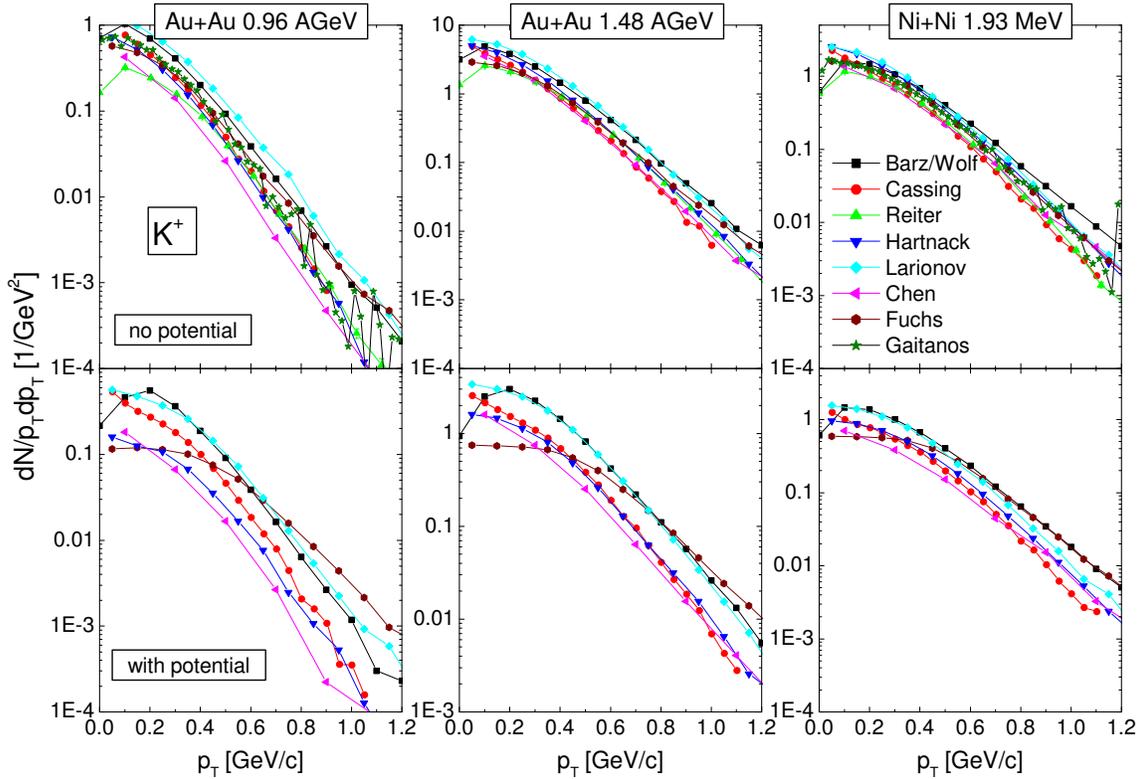}
\caption{Final $K^+$ transverse momentum distribution at $b$=1 fm, $|y_{cm} < 0.5|$ and with an 
enforced $\Delta$ lifetime of 1/120 MeV (top row without, bottom row with KN
potential) in the different approaches.}
\label{kp-pt}
\end{figure}

Fortunately, the uncertainty about the optimal description of the $\Delta$ 
lifetime influences neither the multiplicity nor the form
of the $K^+$ spectra in a significant way 
because low mass $\Delta$'s have a small cross section to produce a $K^+$.
Due to the high threshold in the mesonic channel
the $K^+$ are mostly produced in baryon-baryon collisions. To make comparisons
more straightforward we enforced for the comparison the same
cross sections \cite{tsu} for the $N\Delta\rightarrow K^+$ and the 
$\Delta\Delta\rightarrow K^+$ channels in the different codes. 
These cross sections are dominant 
at $E_{kin} \ge 1.5 AGeV$. Therefore, and due to the constant
$\Delta$ lifetime, the results presented here may differ from already published 
results. 
For the $NN\rightarrow K^+$ cross section we kept the original cross sections of
the programs. For the channel $pp\rightarrow K^+$ new data at low energies 
are available \cite{anke}. In addition there 
is the problem of how to extrapolate this pp cross section to the 
$nn,np\rightarrow K^+$ channels. Usually this extrapolation is done by assuming
that the $K^+$ is accompanied by a $\Lambda$ and that 
only the isospin coefficients change. In the case of the $\kp$ this is,
however, not so easy because this extrapolation depends on whether a kaon or a
pion is exchanged. For the case of a pion exchange we find 
$\sigma(np\rightarrow K^+) = 5/2 \ \sigma(pp\rightarrow K^+)$ and consequently
$\sigma(NN\rightarrow K^+) = 3/2 \ \sigma(pp\rightarrow K^+)$; in the case of a 
kaon we find $\sigma(pp\rightarrow K^+) = 2 \sigma(np\rightarrow K^+)$ 
and therefore $\sigma(NN\rightarrow
K^+) = 1/2 \ \sigma(pp\rightarrow K^+)$. Very recent data \cite{anke1} point,
however, towards
a $\sigma(np\rightarrow K^+)$/$\sigma(pp\rightarrow K^+)$ ratio of 3-4.
A further complication is the production
of the $K^+$ in the $pp\rightarrow K^+N\Sigma$ channel, which has been found to
be small \cite{anke}, but this might be a consequence of the final state interaction.
The different $\sigma(BB\rightarrow K^+)$ cross sections employed
are presented in fig. \ref{kp-cross} and Table \ref{cross}. 
\begin{table}[hbt]
\centering
\begin{tabular}{|c|c|c|c|} \hline
Program  &$\sigma(NN\rightarrow K^+)$ & $\sigma(N\Delta\rightarrow K^+)$ &
$(\sigma
 \Delta\Delta\rightarrow K^+)$ \\ \hline
 Barz  & \cite{tsu}& \cite{tsu}& 0  \\ \hline
 Chen &\cite{tsu} &\cite{tsu} & \cite{tsu}  \\ \hline
 Fuchs &\cite{sib} $(\Lambda)$ \cite{tsu} $\Sigma$& \cite{tsu} & \cite{tsu} \\ \hline
 Gaitanos &\cite{sib} $(\Lambda)$ \cite{tsu} $\Sigma$& \cite{tsu} & \cite{tsu} \\ \hline
 HSD (Cassing) &\cite{cc1} & \cite{tsu}& \cite{tsu} \\ \hline
 IQMD (Hartnack)&\cite{sib} $(\Lambda)$ \cite{tsu} $\Sigma$& \cite{tsu}& \cite{tsu} \\ \hline
 Larionov&\cite{tsu} &\cite{tsu} & \cite{tsu}\\ \hline
\end{tabular}
\caption{The $K^+$ cross sections used in the programs.}
\label{cross}
\end{table}
Top panels of fig. \ref{kp-y} and fig. \ref{kp-pt} present, respectively, 
the $K^+$ rapidity and transverse momentum distribution obtained for those cross
sections.
Also here the URQMD calculations (Reiter) cannot directly be compared with the
others. In this approach, the $K^+$ are created by two body resonance decays
whereas in all other approaches the $K^+$ production is a process with 
three bodies in the final state:
$BB\rightarrow K^+\Lambda (\Sigma)B$.  We see that URQMD is not suited for 1
AGeV but already at 1.5 AGeV it describes the spectra in a reasonable way
although spectra measured in elementary collisions point towards a 
three - body process. In order to understand the 50\% differences 
of the predictions one has to realize that the $K^+$ are produced at subthreshold energies
in the Au+Au collisions. Therefore the nucleons, which are involved 
in the $K^+$ production,
had already collisions before. This encodes in the production yield the whole history
of the reaction. In addition, the $K^+$ production cross section rises exponentially
close to the threshold and therefore small differences in the parametrization of the Fermi
motion of the nucleons or in the potential of the hyperon change the cross section
considerably. Some programs contain, in addition, the channel $\pi B\rightarrow
YK^+$, but usually this channel does not contribute a lot at these energies due
to its large threshold and the few pions as compared to baryons.

Kaons change their properties in the medium. This is the prediction of many
theoretical approaches which range from chiral perturbation theory to
Nambu Jona-Lasinio model calculations~\cite{kpl}. All these theories should give reliable predictions
if the density is low  but suffer from the fact that the extrapolations to 
higher densities are less and less well founded. For our purpose, it is
sufficient to parametrize the density dependence by
\be
  \omega_K(\rho_B,k) = \sqrt{m_K^*(\rho_B)^2 + k^2},
\ee
where the effective mass is
\be
   m_K^*(\rho_B) = m_K^0 \left( 1 - \alpha \frac{\rho_B}{\rho_0} \right),
\label{eq}
\ee
with $\rho_B$ and $\rho_0$ being the baryon density and the normal nuclear matter density,
respectively.
Table \ref{kmass} displays the different values of $\alpha$  employed in the
calculations:
\begin{table}[hbt]
\centering
\begin{tabular}{|c|c|} \hline
Program  &$\alpha [MeV]$ for $K^+ \ (K^-)$\\ \hline
 Barz  &-0.05 (0.16)    \\ \hline
 Chen &-0.04 (0.22)    \\ \hline
 Fuchs &-0.07\\ \hline
 HSD (Cassing) & -0.04 (0.10)    \\ \hline
 IQMD (Hartnack)& -0.075 (0.22)\\ \hline
 Larionov&-0.06 (0.2) \\ \hline
\end{tabular}
\caption{Parameters for the density dependence of the in medium kaon masses $m^*_K$ (eq.\ref{eq}).}
\label{kmass}
\end{table} 
It has been suggested in ref. \cite{rud} to compare the ratio
$\frac{d\sigma^{pC}}{dp_t}/\frac{d\sigma^{pA}}{dp_t}$, which has been measured 
\cite{nek}, with calculations using different $K^+ N$ potentials
\cite{nek}, to fix the $K^+$ potential at normal nuclear matter density.

Despite of the moderate increase of the $\omega_{K^+}(k=0)$ with density, 
the potential changes the $K^+$ yield considerably, as one sees in 
figs. \ref{kp-y} and \ref{kp-pt}. 
This is due to the fact that kaons are produced at high density when 
the mean-free-path 
is short and therefore the $\Delta$'s have a higher chance to collide with a nucleon.
Depending on the density dependence of the potential, we see a decrease of the total $K^+$
yield of about a factor of 2 for the heavy systems and less for the light
system if the potential is switched on. The form of the transverse
momentum distribution by Fuchs differs from those from other calculations 
because he uses a covariant form of nuclear current instead of  only its forth
component, the baryon density,  as the other approaches.

\begin{figure}[hbt]
\includegraphics[width=15cm]{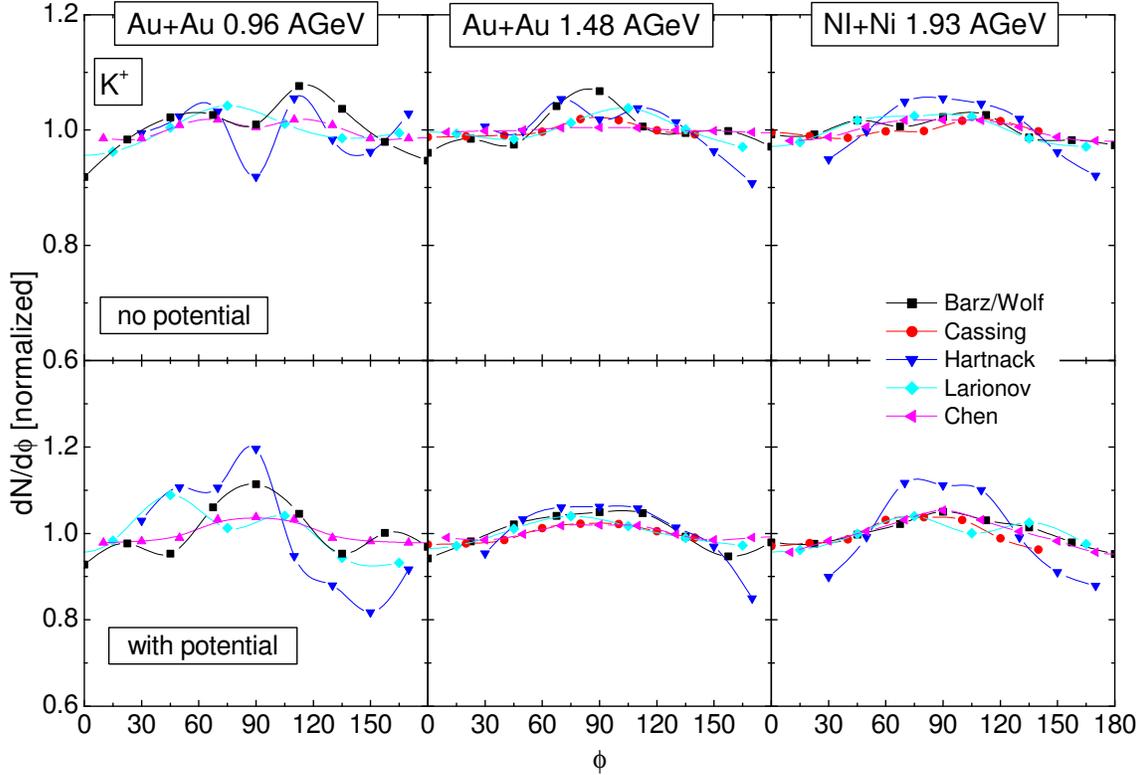}
\caption{Final $K^+$ azimuthal distribution at b=5 fm, $|y_{cm} < 0.5|$ and with an 
enforced $\Delta$ lifetime of 1/120 MeV (top row without, bottom row with KN
potential) in the different approaches.}
\label{kp-phi}
\end{figure}

The elementary $K^+$ creation process is azimuthally isotropic.
It has been argued that the azimuthal distribution of the $K^+$ contains information on the
$K^+$ nucleon potential \cite{bro}. This is based on the fact that $K^+$ emitted into the
azimuthal angle  $\phi = 90^\circ$ come from highly compressed matter whereas those 
being emitted near   $\phi = 0^\circ(180^\circ)$ have traversed the cold spectator matter. 
Therefore one expects
an enhancement around $\phi = 90^\circ$ due to the more repulsive potential. Such an effect
may, however, be superimposed by the consequence of the different path lengths in matter.
Fig.\ref{kp-phi} shows the result. The majority of the programs produce a different azimuthal
distribution without (top row) and with (bottom row) $K^+$N potential. Without the potential,
the distribution is more isotropic and the expected enhancement is seen if the repulsive potential is active.  
\section{$K^-$ Mesons}
\begin{figure}[hbt]
\includegraphics[width=15cm]{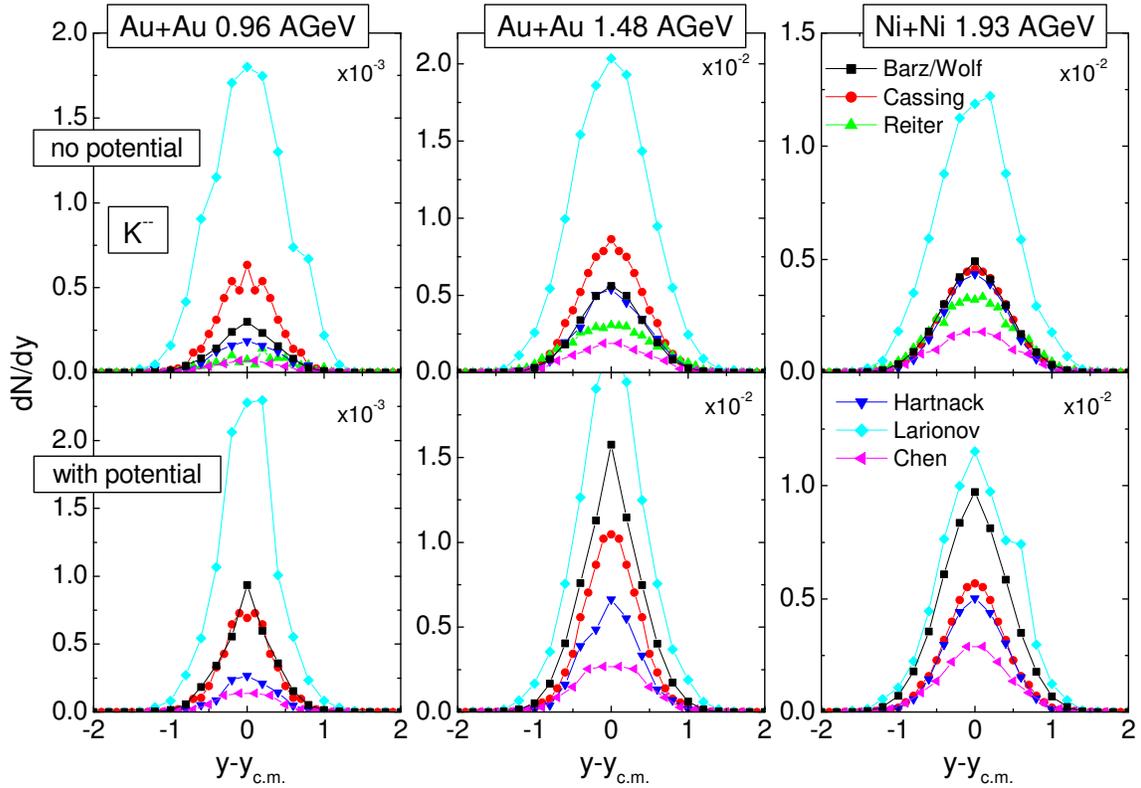}
\caption{Final $K^-$ rapidity distribution at b=1 fm and with an 
enforced $\Delta$ lifetime of 1/120 MeV (top row without, bottom row with KN
potential) in the different approaches.}
\label{km-y}
\end{figure}

\begin{figure}[hbt]
\includegraphics[width=15cm]{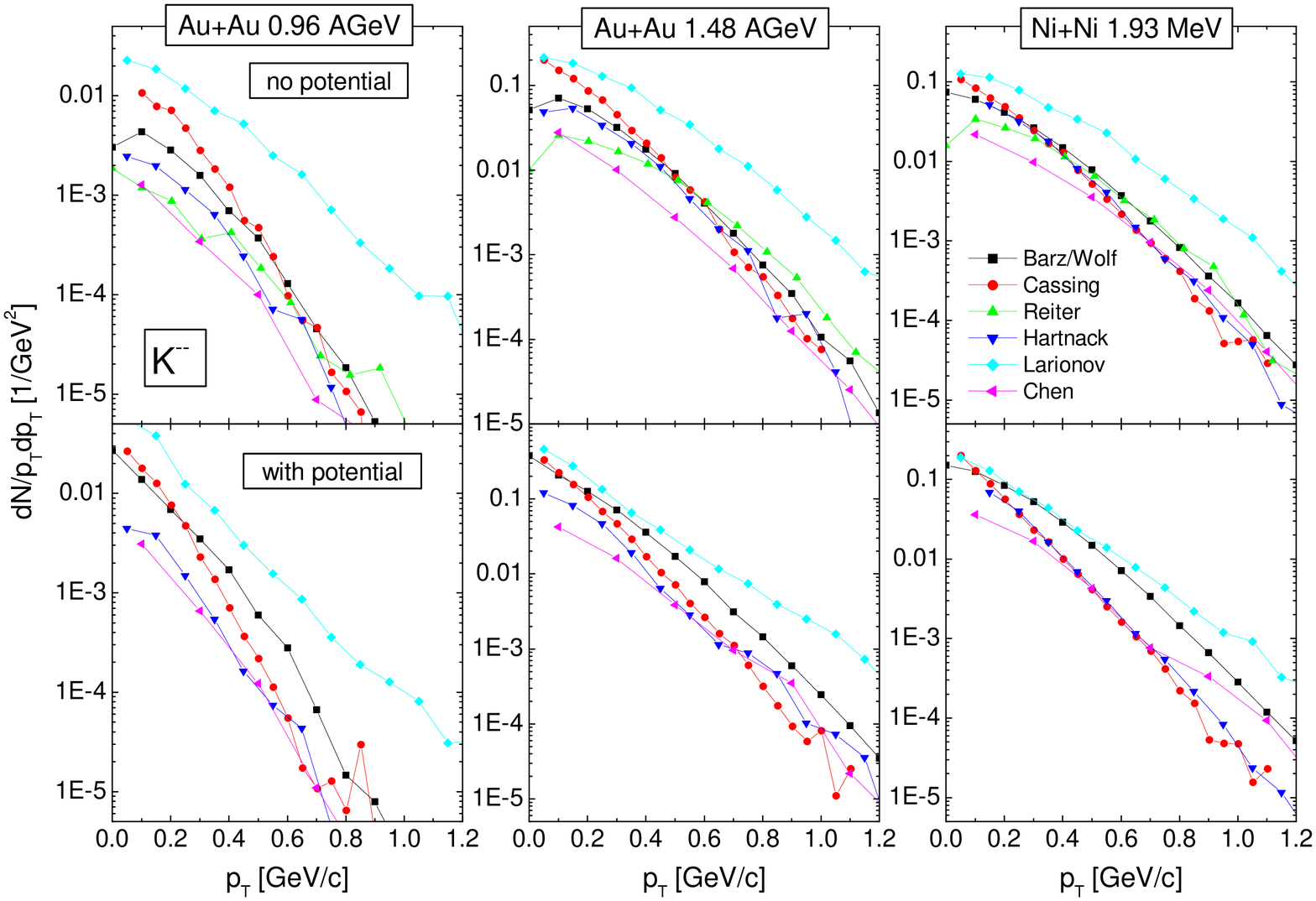}
\caption{Final $K^-$ transverse momentum distribution at b=1 fm, $|y_{cm} < 0.5|$ and with an 
enforced $\Delta$ lifetime of 1/120 MeV  (top row without, bottom row with KN
potential) in the different approaches.}
\label{km-pt}
\end{figure}

The production process for the $K^-$ mesons in nuclear reactions is much more complex
than that for the $K^+$. The reason for this is their s quark content. On  the one hand
it is much harder to produce a $K^-$ in elementary pp reactions because of its higher threshold
($E_{Kin} = 2.513 \ GeV$ for $pp \rightarrow K^+ K^- pp$),  and on the other hand a $K^-$ can exchange its
s quark with strange baryons ($K^- N \rightarrow \Lambda (\Sigma ) \pi$). 
The cross sections for these exothermic reactions are very large. The inverse reaction
$\Lambda (\Sigma ) \pi \rightarrow K^- N$ has been identified as the dominant
source of the $K^-$ production in heavy ion reactions at the energies
we consider here \cite{ko1}. 
This process is absent in elementary pp interactions and therefore the $K^-$ yield in
heavy ion collisions is orders of magnitudes larger than that in pp collisions for energies not
too far away from the threshold. This process couples the $K^-$ to the $\Lambda (\Sigma)$ 
and hence to the $K^+$ production~\cite{kvk96,hartprl,cley} and explains therefore the experimental
observation \cite{foe} that the $K^+$ to $K^-$ ratio is independent of the impact parameter
The $K^-$ production in baryon-baryon interactions suffer from the large threshold 
and takes place only early when the system is dense. Thereafter the $K^-$ gets easily reabsorbed.
In figs. \ref{km-y} and \ref{km-pt} we display the $K^-$ rapidity and transverse momentum
distributions. 
The URQMD calculations suffer from the fact that they produce too few strange
baryons (produced in the $ BB \rightarrow K^+ Y$ reactions) as can been seen from the
low $K^+$ multiplicity. 
The $K^-$ channel confronts the simulation programs with the challenge to
calculate the $K^-N$ and the $\Lambda \pi$ cross sections at values 
of $\sqrt{s}$ which are in free space below the threshold. These problem has
been solved differently by the different groups. 

In view of the complexity of the production processes, the results of those codes, which can be
directly compared, are rather similar, especially at higher energies.
At higher energies, the transverse momentum distributions are also rather similar.  
At lower energies the HSD code produces a somehow softer distribution.
Also the $K^-$ changes its properties in the medium. As pointed in \cite{kvk95}, 
due to the strange resonances this change is rather complex.In fact the detailed
calculations \cite{lut}, based on the well constrained $\bar K N$ interaction from
\cite{lk},
raise the question whether the $K^-$ can still be treated
as a quasi particle at high densities and temperatures.
On the other hand, most of the observed
$K^-$ are produced not too far away from the surface and therefore the multiplicity should not
be too much influenced by the kaon properties at high density. Currently, one has approached this problem 
by using the same
form of the mean field potential as for the $K^+$ 
with the parameters displayed in table \ref{kmass}. The energy $\omega _{K^-}(k=0)$
decreases with increasing density and therefore one expects an increase of the $K^-$ yield if
the potential is switched on. This is indeed observed in the bottom figures of the figs.\ref{km-y} and \ref{km-pt}.
The enhancement seen when the $K^-$ potential is switched on is different in the
different approaches. An important part of this difference comes from
different extrapolations of the known $K^-N$ cross section, 
which diverges at small relative momenta to the $\sqrt{s} < \sqrt{s_{threshold}}$ region. 
A solid theoretical basis for this extrapolation, which is necessary because the mass of the $K^-$
decreases and hence the threshold is lowered, is not at hand presently.   
 
\section{Lasting results}
One of the most challenging motivations for studying heavy ion collisions has been the search for the 
amount of energy which is needed to compress hadronic matter. This went under the slogan "Search of
the nuclear equation of state". Earlier ideas to use pions or the in-plane flow of baryons
as an experimental signal for the compressibility of matter, have not led to unambiguous
results. The pions turned out to be not sensitive because most of them are produced when
the system has already expanded to a low density. The in-plane flow is small at densities
reachable in these reactions and it depends on the pressure which is proportional to the density 
gradient. Nuclear surface properties, like density gradients, are in general much more 
difficult to simulate than bulk properties. 

\begin{figure}[hbt]
\includegraphics[width=15cm]{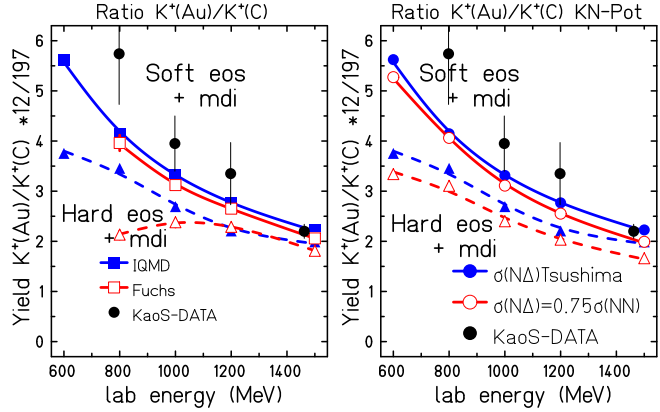}\vspace*{-12cm}
\caption{
Normalized ratio of the $K^+$ yield in Au+Au to C+C reactions as a function
the beam energy for two different equations of states. Filled circles with errors
represent the data of the KaoS collaboration \cite{sturm}. Other symbols with lines represent
calculations of ref. \cite{fuc} on the left panel and of ref. \cite{iqmd} on the right panel
for either the $\Delta N \rightarrow K^+$ cross section of ref.\cite{tsu} or for
($\sigma(\Delta N\rightarrow K^+) = 3/4 \ \sigma(NN\rightarrow K^+)$) employed earlier.
}
\label{eos}
\end{figure}

It has been proposed to
use $K^+$ for this purpose because at these subthreshold energies they test the nuclear
environment prior to production \cite{ko}. To get rid of many of the uncertainties it has been proposed
by the KaoS collaboration \cite{sturm} to use not the absolute yield but to compare a heavy with a light system where 
compression is virtually absent. Indeed, one has found that the excitation function of this
ratio is rather sensitive to the compressibility of the nuclear potential and that the
experimental results point strongly towards a nuclear potential with a compressibility $\kappa
\le 280 MeV$. Such a potential has usually been dubbed as the "soft equation of state". In the meantime,
one has tested whether this result is robust, i.e. whether it remains valid if one varies
the unknown or little known input such as the $\Delta$ lifetime or the $\Delta N \rightarrow
K^+ $ cross section with the conclusion that only a strongly density dependent $K^+$ production
cross section may invalidate this result. Such a cross section, however, has never been
proposed in elementary calculations. To which degree this result is stable against 
a change of the $\Delta N \rightarrow K^+ $ cross section is shown in the left panel of 
fig. \ref{eos} where the different cross sections vary by a factor of 2.
The results appear fairly robust. One should stress, though, that we find here that
nuclear matter is relatively easy to compress. The compressibility is compatible with the
results obtained by analyzing giant resonances, i.e. the compression of nuclei at 
very low temperature and around the equilibrium density. However, the reactions discussed here
yield high densities and temperatures and the system contains mesons and resonances. Thus we test here a
completely different region in the T-$\rho$ plane. 
\section{Conclusion} 
Multiplicity, rapidity and transverse momentum distributions from different programs
simulating heavy ion reactions at kinetic energies of around
1 AGeV have been compared. If the same inputs are used, the programs 
yield quite similar results but differ in details. Some of the differences are understood
in terms of the implementation of different physical approaches in the programs, 
and others, less important, remain
to be explained. Despite of these differences the programs agree on some important physical
results: The pions are almost exclusively produced by resonance decay. 
Subthreshold meson production is due to the same reaction as in free space
and becomes possible because at least one of the collision partners has 
gained energy in collisions before. They programs furthermore agree on the 
stopping of matter, the dependence of the $K^+$ squeeze out on its in-medium potential and 
on the softness of the density dependence of the effective nuclear potential.

{\it Acknowledgment} We would like to thank Dr. H. Oeschler for many valuable 
discussions and the
$ECT^*$ in Trento, Italy, for hosting a workshop which was the starting point of this comparison.

\end{document}